\documentstyle[11pt,aaspp4,psfig,lscape]{article}
\doublespace

\begin{document}

\title{Sulphur abundance in very metal-poor stars\footnote{
Based on observations  obtained with the William Herschel
Telescope, operated on the island  of La Palma by the Isaac Newton 
Group in the Spanish Observatorio  del Roque de los Muchachos 
of the Instituto de Astrof\'\i sica de  Canarias.}}

\author{Garik~Israelian and Rafael~Rebolo\altaffilmark{2}}
\affil{Instituto de Astrof\'\i sica de Canarias, E-38200
La Laguna, Tenerife, Spain} 

\authoraddr{Instituto de Astrof\'\i sica de Canarias, E-38200 La Laguna, 
Tenerife, Spain}

\altaffiltext{2}{Consejo Superior de Investigaciones Cient\'\i fcas, Spain}

\authoremail{gil@ll.iac.es, rrl@ll.iac.es}

\begin{abstract}

We have obtained high-resolution high signal-to-noise spectra of the S\,{\sc i}
near-infrared doublet at 8694 \AA\ in eight  metal-poor stars with metallicities
in the range  $-0.6\leq$[Fe/H]$\leq -3.0$. Elemental sulphur abundances were
derived for six targets and upper limits were set for two of the most metal-poor
stars in the sample. The sulphur doublet at 8694.62 \AA\ has been detected in 
three stars (HD\,19445, HD\,2665 and HD\,2796)  with [Fe/H]$\leq -1.9$.  
Our observations combined with those available in the literature indicate a monotonic
increase of the  [S/Fe] ratio as [Fe/H] decreases, reaching values [S/Fe]$\sim$0.7-0.8
below [Fe/H]=$-$2. We discuss plausible scenarios for interpretation of these
results.

\end{abstract}


\section{Introduction}

The standard scenario for the chemical evolution of the Galactic Halo
(see e.g. Tinsley 1979) is based on the different  yields of Type Ia and
Type II Supernovae. It is well established that Type II supernovae mostly
produce the so-called $\alpha$-elements (O, Ca, Mg, S, Si and Ti) while 
Type Ia supernovae are responsible for the synthesis of the Fe-group
(Fe, Co, Ni, Zn). Given a different life-times of both types of supernovae
(10$^7$ years for Type II and 10$^9$ years for Type Ia), it is expected
that there will be overabundance of $\alpha$ elements  at [Fe/H]$< -1$. 
At first glance, this behavior is observed in many metal-poor halo star
(McWilliam 1997). The precise trends of the [$\alpha$/Fe] ratio versus 
[Fe/H] are still not clearly established. For instance, low [$\alpha$/Fe] ratios 
were found in several metal-poor halo stars (Carney et al. 1997) and a rather
complicated picture emerged from the comprehensive work by Idiart \& Thevenin (2000)
who studied  [Ca/Fe] and [Mg/Fe] ratios in  252 halo and disk stars. 
On the other hand, the [O/Fe] ratio in  unevolved stars show a linear rise 
(Israelian, Garc\'\i a L\'opez \& Rebolo 1998, Israelian et al. 2001)
with a slope $-$0.33$\pm$0.02 in the metallicity range 0$\leq$[Fe/H]$\leq -3$. 
Apparently there is no general agreement about a unique $plateau$ 
[$\alpha$/Fe]$\approx const$ at [Fe/H]$< -$1 for  $\alpha$-elements 
such O, Ca and Mg. This statement can be extended to another $\alpha$-element:
sulphur. 

In his extensive analysis Fran\c{c}ois (1987, 1988) was able to delineate the [S/Fe]
trend down to the metallicity [Fe/H]=$-$1.6 suggesting that 
the [S/Fe] ratio in the halo could be constant at a value around 0.6. The goal
of this letter is to investigate whether unevolved halo stars  at lower 
metallicities support the proposed trend.

\section{The Observations}

The observations were carried out in 2000 August 17-19 using the Utrecht 
Echelle Spectrograph (UES) at the Nasmyth focus of the 4.2-m WHT at the ORM 
(La Palma). The CCD detector was a 2148$\times$2148 pixel$^2$
SITE1. We obtained echelle spectra with a slit width $\sim$1\farcs{0} which 
cover the wavelength range between 5220 and 9150 \AA\ in 20 orders at a 
spectral resolving power of $R=\lambda /\Delta\lambda\sim 50,000$. 
All spectra were background and scattered light subtracted, flat-fielded
and extracted using standard {\sc iraf} \footnote{{\sc iraf} is
distributed by the National Optical Astronomical Observatories, which is
operated by the Association of Universities for Research in Astronomy, Inc.,
under contract with the National Science Foundation, USA.} procedures.
The flat-field correction was attempted with a fast rotator, tungsten
lamp and dome flats. However, the best correction from a severe fringing
in the near-infrared was achieved with high quality dome flats. 
The wavelength calibration was performed with a Th--Ar lamp giving a 
dispersion of 0.087 \AA\/pixel.
The final signal-to-noise ratio of our spectra is provided in Table 1.

\section{Spectral synthesis, stellar parameters and abundances}

The LTE spectrum synthesis code MOOG (Sneden 1973) and ATLAS9 model
atmospheres (Kurucz 1992) were used in the present study. Atomic parameters 
of the sulphur doublet 8693.93 \AA\ ($\log gf$=$-$0.5, $\chi$=7.87 eV) 
and 8694.62 \AA\ ($\log gf$=0.08, $\chi$=7.87 eV)  
were obtained from the VALD-2 database (Kupka et al. 1999). 
These $\log gf$-values combined with the equivalent widths  
28 and 11 m\AA\ of the sulphur doublet derived from the high-resolution solar
atlas (Kurucz et al. 1984) provide a solar sulphur abundance 
$\log \epsilon$(S)=7.20$\pm$0.02 (on the customary scale in which 
$\log \epsilon$(H)=12). This is in excellent agreement with 
$\log \epsilon$(S)=7.21 given by Anders \& Grevesse (1989). 

All our targets (Table 1) are well known metal-poor stars. The IRFM based 
effective temperature, $T_{\rm eff}$, of two giants HD\,2796, BD$-$18\arcdeg5550 
and a subgiant HD\,2665 were taken from Alonso, Arribas,  \& Mart\'\i nez-Roger (1999) 
while the surface gravities and [Fe/H] values were estimated from Fe NLTE
analysis (Israelian et al., in preparation) using the same tools as 
in Israelian et al. (2001). In any case, the parameters of these three
evolved stars  (listed in Table 1) are very similar to those reported in the
literature (Pilachowski, Sneden \& Booth 1993, Alonso et al. 1999, 
Burris et al. 2000). The $T_{\rm eff}$ of five subdwarfs (Table 1) are also based 
on the IRFM and come from Israelian et al. (1998). Their metallicities and surface 
gravities were adopted from the NLTE Fe study of Thevenin \& Idiart (1999) and 
agree perfectly with those estimated using  {\it Hipparcos} parallaxes 
(Allende Prieto et al. 1999). We note that our $T_{\rm eff}$-values agree 
with those of Thevenin \& Idiart (1999) within 60 K.   
NLTE effects on sulphur are negligible (less than 0.05 dex) due to the weakness
of the sulphur lines in our targets (Takada-Hidai et al. 2001).

Our tests show that a typical change of +100 K in $T_{\rm eff}$ increases
[S/H] by $-$0.05 and $-$0.09 dex for $\log g$=4.15 and $\log g$=1.5, respectively.
A change of surface gravity by +0.3 dex implies an uncertainty 0.11 and 0.13 dex
for $T_{\rm eff}$=5800 and 4900 K, respectively. The errors arising from
the uncertainties in the metallicity and the microturbulent velocity $V_{mic}$
are negligible. We set $V_{mic}$=2\,\hbox{km$\;$s$^{-1}$} in HD\,2796, HD\,2665,
BD$-$18\arcdeg5550 and $V_{mic}$=1\,\hbox{km$\;$s$^{-1}$} for the dwarfs. The sulphur
abundance for each star (Table 1) was obtained as the weighted mean of the 
abundances resulting for each line of the doublet (with weights proportional
to the equivalent widths). The final error in the sulphur abundance was
 derived as a quadratic sum of the errors associated with the stellar parameters and
the errors associated to the equivalent width measurements. The final errors for
the most metal-rich five dwarfs are similar to those reported by Fran\c{c}ois (1987, 1988)
 being  of the order of $\pm$0.1 dex. The errors are larger in the most metal-poor 
targets of the sample. 

We have reanalyzed sulphur abundances from Fran\c{c}ois (1987, 1988) using 
the equivalent widths provided in these papers. Effective temperatures (Table 2) 
have been estimated using the IRFM calibration of Alonso, Arribas,  \& Mart\'\i nez-Roger 
(1996) for Str\"omgren $b-y$ and c$_1$ indices taken from Hauck and Mermillod 
(1998) and [Fe/H] adopted from Thevenin \& Idiart (1999), Allende Prieto et al. (1999) 
and Allende Prieto \& Lambert (1999). No reddening corrections were applied since all 
these stars are nearby within 100 parsec (Perryman et al. 1997). Typical errors 
in $T_{\rm eff}$ are $\pm$100 K.  Allende Prieto et al. (1999) have shown that 
gravities derived from Fe NLTE analysis by Thevenin \& Idiart (1999) agree with 
the gravities inferred from accurate {\it Hipparcos} parallaxes in the metallicity 
range $-1 \leq [Fe/H] \leq 0$. 
Surface gravities of 
these stars were estimated using the {\it Hipparcos} parallaxes. We used $V$ 
magnitudes from {\it Hipparcos} Catalogue to compute $M_{\rm v}$  and 
to estimate absolute bolometric magnitudes $M_{\rm bol}$ from Alonso, 
Arribas, \& Mart\'\i nez-Roger (1995). Therefore the set of new parameters 
listed in Table 2 is homogeneous (three stars have not been considered
because their parameters were rather uncertain). The masses of stars were 
inferred using theoretical isochrones from evolutionary tracks of 
the Geneva group (Schaerer et al. 1993). Our gravities have typical errors 
$\pm$0.2 dex and are in good agreement with those presented by different 
authors (Allende Prieto et al. 1999, Allende Prieto \& Lambert 1999) using, 
however, different evolutionary tracks.

\section{Discussion and Conclusions}

The [S/Fe] ratios obtained for our targets are listed in Table 1 and plotted in 
Fig. 3 together with Fran\c{c}ois (1987, 1988) measurements.  NLTE estimates for 
the metallicities of his stars  have been made using the polynomial fit from
Israelian et al. (2001) which is based on the work of Thevenin \& Idiart (1999). 
Our results are consistent with those by Fran\c{c}ois in the overlapping metallicity 
range. Remarkably, at very low metallicities we find the highest [S/Fe] ratios supporting
a progressive increase of the [S/Fe] ratio as metallicity decreases. A minimum-square
linear fit for the points in Fig. 3 results in a slope of $-$0.46 $\pm$0.06 (1-$\sigma$) 
which can be compared with the slope in the [O/Fe] vs. [Fe/H] relationship of 
$-$0.33  $\pm$0.02 (1-$\sigma$) found by Israelian et al. (2001).
 
Sulphur yields from standard Type II supernovae have been computed by many
authors  (e.g., Woosley \& Weaver 1995; Thielemann, Nomoto, \& Hashimoto 1996). 
The [S/Fe] ratio in the ejecta is sensitive to the amount of mass going into the
compact object formed during the supernova explosion (i.e depend on the adopted
mass cut). There is a range of
values for this mass cut  which can perfectly provide [S/Fe] yields consistent
with the observations. The tentative linear trend delineated in Fig. 3 may 
be an indication that the explosive energy of  Type II supernovae may depend
on progenitor mass. Alternatively,  Nomoto et al. (2001) and Nakamura et al. (2001) 
have presented models of explosive nucleosynthesis in hypernovae. They 
consider exploding He cores with different initial masses and explosion
energies in the range 1--100 $\times$ 10$^{51}$ erg. Their computations show
that [S/Fe] ratios as those found in very low-metallicity stars can easily be
produced in these explosions. For instance, it can be seen from Table 2
of Nakamura et al. (2001) that the He core with a mass 16~M$_{\sun}$  and 
10$\times$10$^{51}$ erg explosion energy is able to provide a ratio [S/Fe]$\sim$1.1 
in the final ejecta. It is possible that the linear trend of [S/Fe] can be explained 
by an increasing role of hypernovae in the early epochs of the Galaxy formation 
(i.e., [Fe/H] $< -1$). The estimated Galactic rate of Type Ic hypernovae is 
$10^{-3}$~yr$^{-1}$ (Hansen 1999) but it could have been  much higher in the
early Galaxy. Small mass-loss rates due to the low metallicity in the first
generations of massive halo stars could have led to large helium cores and
the conservation of the primordial angular momentum. Since hypernovae
are favored by stars with a large helium core mass and rapid rotation,
a significant sulphur production in the early Galaxy could have
been a rather common phenomenon. Similar ideas have been put forward by 
Qian and Wasserburg (2001) to explain a linear rise in the [O/Fe] ratio.
Extension of their model for the sulphur case would be very interesting.

Additional observational support for  large amounts of sulphur synthesized 
during Type II supernova/hypernova explosion  comes from the analysis of
the chemical composition of the  low-mass  star orbiting the black hole 
in the X-ray binary system  GRO J1655-40 (Israelian et al. 1999). 
 $\alpha$ elements are uniformly enhanced in the atmosphere of the secondary star, 
and in particular a  large amount of sulphur [S/H]$\sim$0.9  was found in the
secondary star. Most probably, this is a result of contamination from the matter 
ejected after supernova or hypernova explosion of the black hole stellar progenitor 
(Israelian et al. 1999, Brown et al. 2000). Very recently, Combi et al. (2001)
have detected a bubble in the large-scale HI distribution around GRO J1655-40
providing new evidence for a supernova explosion in this system.

Another possible explanation for the monotonic increase in [S/Fe] at low [Fe/H]
is related to a delayed deposition of the supernova synthesized products into
the ISM due to differences in transport and mixing. Choosing a short mixing 
time ($\sim$ 1\,Myr) for oxygen (a volatile), and longer mixing time
($\sim$ 30\,Myr) for Fe (refractory), Ramaty et al. (2000, 2001) suggest that 
[O/Fe] should rise monotonically with a slope which is consistent with the 
observations of Israelian et al. (1998, 2001). A similar trend is also 
predicted (Ramaty et al. 2001) for a volatile element like sulphur. 
As dicussed by Ramaty et al. (2000, 2001), it is not necessary that other
$\alpha$-elements like Mg and Si follow the same trend as O and S since 
they are refractory elements with much longer mixing time in  the ISM.      
In summary, while more observations are clearly needed at very low metallicities
in order to confirm the trend delineated in the present work, our new measurements
do not seem to support a [S/Fe] $plateau$ versus [Fe/H] in halo stars. There are
several scenarios that can account for a monotonic increase in [S/Fe]
at low metallicities.

\vspace{0.1cm}

\begin{acknowledgements} 
We thank Y. Takeda and M. Takada-Hiadi for providing the NLTE corrections 
for sulphur IR doublet before the publication. This work was partially
funded by project DGES PB 98-0531-c02-02.

\end{acknowledgements}

\clearpage

\begin{landscape}
\begin{deluxetable}{lcccccccccccc}
\scriptsize
\tablecaption{Observing log, stellar parameters and sulphur abundances in the sample of metal-poor stars.
The 1-$\sigma$ errors in the equivalent widths are computed with the formula by 
Cayrel (1988). The
upper limits for HD\,140283 and BD-18\arcdeg555 correspond to a 3-$\sigma$ values.
\label{tbl-1}}
\tablewidth{0pt}
\tablehead{
\colhead{Target} &
\colhead{$V$}    &
\colhead{Exposure time (s)} &
\colhead{S/N}           &
\colhead{EW (m\AA)} &
\colhead{EW (m\AA)} &
\colhead{$T_{\rm eff}$}  &
\colhead{$\log g$}       &
\colhead{[Fe/H]}    &
\colhead{[S/H]}          &
\colhead{[S/Fe]}         &
\colhead{Ref.}           \nl
& & & & 8694.6 \AA\ & 8693.9 \AA\ &  &  &  &  &  &  \nl
}
\startdata
HD\,2665   &  7.75 & 3200    & 400   & 3.0$\pm$0.6   &  -  & 4990$\pm$66 & 2.50$\pm$0.2  & $-$2.0  & $-$1.31$^{+0.13}_{-0.14}$ & 0.69 & 1\nl
HD\,2796   &  8.51 & 5000    & 350   & 2.6$\pm$0.7   &  -  & 4867$\pm$58 & 1.80$\pm$0.2  & $-$2.3  & $-$1.49$^{+0.17}_{-0.18}$ & 0.81 & 1\nl
HD\,19445  &  8.05 & 10\,900 & 650   & 2.7$\pm$0.4 &  -  & 5810$\pm$150 & 4.46$\pm$0.2   & $-$1.88 & $-$1.23$^{+0.11}_{-0.12}$ & 0.65 & 2\nl
HD\,140283 &  7.24 & 2500    & 300   & $\leq$2.5   &  -  & 5550$\pm$100 & 3.74$\pm$0.2   & $-$2.21 & $\leq -1.23$ & $\leq$0.98 & 2 \nl
HD\,157214 &  5.40 & 150     & 350   & 20$\pm$0.7    & 8.0$\pm$0.7 & 5625$\pm$100 & 4.36$\pm$0.2   & $-$0.32 & $-$0.13$\pm$0.09  & 0.19 & 2 \nl
HD\,194598 &  8.36 & 5400    & 350   & 9.0$\pm$0.7   & 3.0$\pm$0.7 & 5960$\pm$120 & 4.38$\pm$0.2   & $-$1.06 & $-$0.77$\pm$0.1  & 0.29  & 2 \nl
HD\,201889 &  8.04 & 1500    & 250   & 14$\pm$1    & 5.0$\pm$1 & 5615$\pm$100 & 4.24$\pm$0.2      & $-$0.94 & $-$0.37$\pm$0.1  & 0.57 & 2 \nl
BD-18\arcdeg555 & 9.34 & 5000    & 300   & $\leq$2.5      &  -   & 4668$\pm$63 & 1.5$\pm$0.3       & $-$2.7  & $\leq -1.3$    &  $\leq$1.4 & 1   \nl
\enddata
\tablerefs{
(1): Alonso et al. (1999), Israelian et al., in preparation,
(2): Thevenin and Idiart (1999), Israelian et al. (1998).
}
\tablenotetext{a}{The  signal-to-noise ratio (S/N) is provided at 8697 \AA.}
\end{deluxetable}
\end{landscape}

\clearpage

\begin{deluxetable}{lcccccc}
\scriptsize
\tablecaption{Reanalysis of sulphur abundances from the samples of 
Fran\c{c}ois (1987, 1988). The metallicities were adopted from the
references listed in the last column. Typical error in [S/Fe] is of
the order of $\pm$0.1 dex.\label{tbl-3}}
\tablewidth{0pt}
\tablehead{
\colhead{Star}           &
\colhead{$T_{\rm eff}$}  &
\colhead{$\log g$}       &
\colhead{[Fe/H]}         &
\colhead{[S/H]}          &
\colhead{[S/Fe]}         &
\colhead{Ref.}           \nl
}
\startdata
HD\,24616   & 4943 & 3.04   & $-$0.88  &  $-$0.51 &    0.37  & 3  \nl
HD\,59984   & 5916 & 3.90   & $-$0.52  &  $-$0.70 & $-$0.18  & 1  \nl
HD\,63077   & 5752 & 4.17   & $-$0.53  &  $-$0.55 & $-$0.02  & 1  \nl
HD\,69897   & 6263 & 4.25   & $-$0.12  &  $-$0.41 & $-$0.29  & 1  \nl
HD\,76932   & 5857 & 4.08   & $-$0.79  &  $-$0.64 &    0.15  & 1  \nl
HD\,88218   & 5677 & 3.87   & $-$0.42  &  $-$0.29 &    0.13  & 3  \nl
HD\,91324   & 6181 & 3.98   & $-$0.23  &  $-$0.52 & $-$0.29  & 2  \nl
HD\,94028   & 5923 & 4.28   & $-$1.31  &  $-$1.12 &    0.19  & 1  \nl
HD\,102365  & 5603 & 4.40   & $-$0.08  &  $-$0.41 & $-$0.33  & 2  \nl
HD\,106516  & 6153 & 4.30   & $-$0.66  &  $-$0.56 &    0.1   & 2  \nl
HD\,121384  & 5185 & 3.39   & $-$0.34  &  $-$0.79 & $-$0.35  & 2  \nl
HD\,132475  & 5619 & 3.79   & $-$1.37  &  $-$1.13 &    0.24  & 1  \nl
HD\,136352  & 5637 & 4.31   & $-$0.21  &  $-$0.47 & $-$0.26  & 2  \nl
HD\,148816  & 5831 & 4.12   & $-$0.51  &  $-$0.55 & $-$0.04  & 1  \nl
HD\,157089  & 5770 & 4.04   & $-$0.39  &  $-$0.50 & $-$0.11  & 1  \nl
HD\,188376  & 5369 & 3.62   & $-$0.08  &  $-$0.30 & $-$0.22  & 3  \nl
HD\,193901  & 5668 & 4.50   & $-$0.97  &  $-$0.70 &    0.13  & 1  \nl
HD\,201891  & 5841 & 4.25   & $-$0.87  &  $-$0.88 & $-$0.01  & 1  \nl
HD\,203608  & 6067 & 4.29   & $-$0.57  &  $-$0.69 & $-$0.12  & 1  \nl
HD\,211998  & 5271 & 3.36   & $-$1.25  &  $-$1.40 & $-$0.15  & 1  \nl
\enddata
\tablerefs{
(1): Thevenin \& Idiart (1999),
(2): Allende Prieto et al. (1999),
(3): Allende Prieto \& Lambert (1999).
}
\end{deluxetable}

\clearpage

\begin{figure}
\psfig{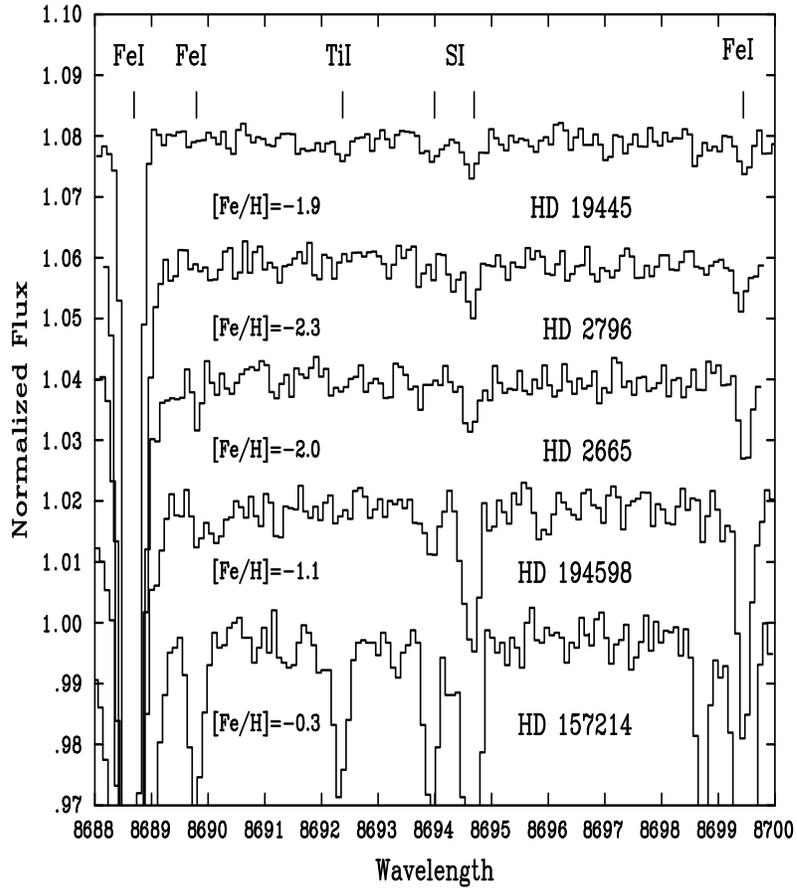}
\caption[]{Sulphur lines in the UES spectra of five metal-poor stars.}
\end{figure}

\clearpage

\begin{figure}
\psfig{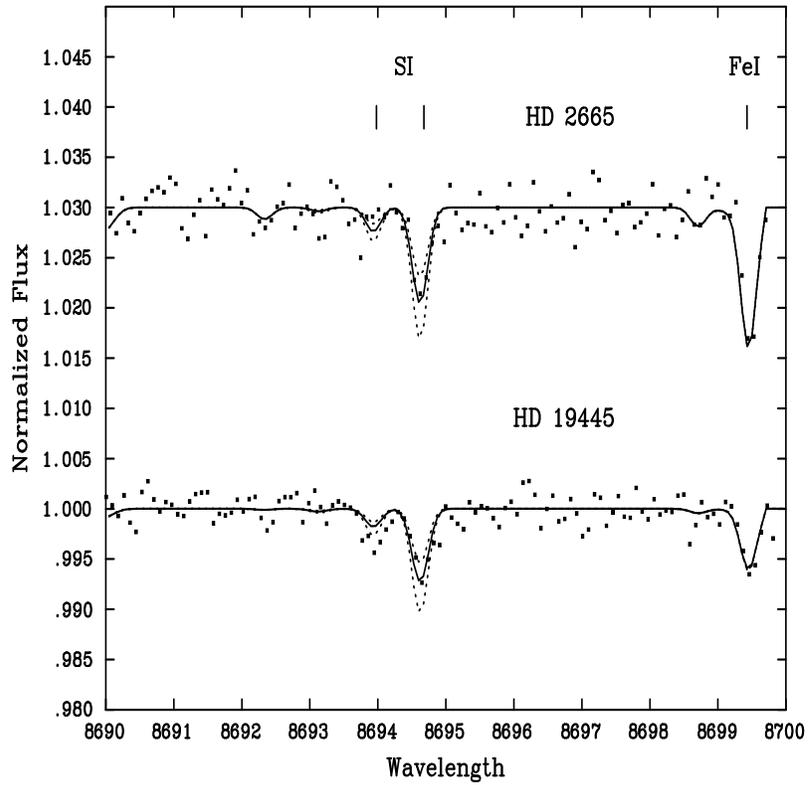}
\caption[]{Synthetic spectra fits to the sulphur lines in the UES 
spectra (dots) of HD\,19445  and HD\,2665. The solid lines are
best fits to the data corresponding to [S/Fe]=0.65 and 0.7 in 
HD\,19445 and HD\,2665, respectively. Thin dotted lines indicate
spectral synthesis with [S/Fe] shifted by $\pm$0.15 dex with respect
the previous best fit values in each star. A vertical shift of 0.03
has been applied to the spectra of HD 2665 for clarity.}
\end{figure}

\clearpage

\begin{figure}
\psfig{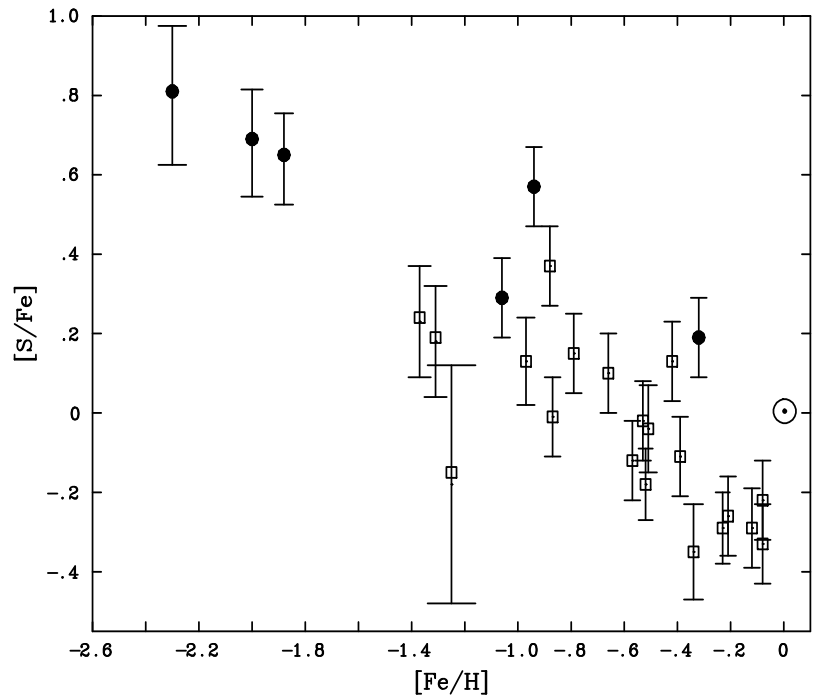}
\caption[]{Sulphur abundances with respect to iron, derived from 
the near-IR S\,{\sc i} lines. Unfilled squares are the data from 
Fran\c{c}ois (1987, 1988) and filled circles are new measurements.
The sun is marked with the standard symbol. Upper limits for two 
very metal-poor stars are not plotted.}
\end{figure}

\end{document}